\documentclass[prl,twocolumn,showpacs]{revtex4}

\usepackage{graphicx}
\usepackage{dcolumn}
\usepackage{bm}

\begin{document}


\title{Dynamically dominant excitations of string solutions in the spin-1/2 antiferromagnetic Heisenberg chain in magnetic fields}

\author{Masanori Kohno}
\affiliation{WPI Center for Materials Nanoarchitectonics, 
National Institute for Materials Science, Tsukuba 305-0047, Japan}

\date{\today}

\begin{abstract}
Using Bethe-ansatz solutions, we uncover a well-defined continuum in dynamical structure factor 
$S^{+-}(k,\omega)$ of the spin-1/2 antiferromagnetic Heisenberg chain in magnetic fields. 
It comes from string solutions which continuously connect 
the mode of the lowest-energy excitations in the zero-field limit and 
that of bound states of overturned spins from the ferromagnetic state near the saturation field. 
We confirm the relevance to real materials through comparisons with experimental results.
\end{abstract}


\maketitle
The spin-1/2 antiferromagnetic Heisenberg chain exhibits intriguing quantum many-body effects, 
associated with modern concepts in condensed-matter physics, such as 
spin liquids, quantum criticality and fractionalization. 
Also, this system is an excellent platform to make precise comparisons between experiments 
and theories: Various intriguing features predicted by exact solutions~\cite{BetheAnsatz} 
can be confirmed by accurate experiments on quasi-one-dimensional materials. 
Actually, it is established that dynamical properties in the absence of magnetic field are 
characterized by quasiparticles called spinons~\cite{FT,HSspinon} through precise comparisons 
between theoretical predictions and experimental results~\cite{KCuF3,CuBenz,CuPzN0}. 
\par
In magnetic fields, dynamical properties are more complicated. 
Some basic features of dominant excitation spectra can be understood by modifying 
the 2-spinon continuum in zero field according to the strength of magnetic fields 
in the $k$-$\omega$ plane \cite{Ishimaru,Muller,CuPzN}. 
The distributions of spectral weights in $S^{-+}(k,\omega)$ and 
$S^{zz}(k,\omega)$ are effectively expressed by low-order particle-hole excitations 
in Bethe-ansatz solutions \cite{Muller,Karbach_Szz,Karbach_psinon}: 
The dominant excitations in $S^{-+}(k,\omega)$ and $S^{zz}(k,\omega)$ 
are known as 2-psinon (2$\psi$) excitations and psinon-antipsinon ($\psi\psi^*$) 
excitations, respectively \cite{Karbach_Szz,Karbach_psinon}. Their properties 
have been investigated in detail by using Bethe-ansatz 
solutions \cite{Karbach_Szz,Karbach_psinon,Fledderjohann,CauxPRL,CauxAffleck,BiegelXXZ,Sato,Caux,Biegel,Nishimoto}. 
\par
As for $S^{+-}(k,\omega)$, the situation is much more complicated. Except the low-energy 
modes near momentum $k$=0 \cite{Muller,Karbach_psinon,Biegel,Nishimoto,Lefmann} and 
$k$=$\pi$ \cite{Muller,Karbach_psinon,Nishimoto,Lefmann}, behaviors 
of $S^{+-}(k,\omega)$ have not been clarified: 
In Ref.~\cite{Muller}, the continuum of $\psi\psi^*$ excitations was predicted 
to persist in the thermodynamic limit based on the classification by the Bethe formalism. 
However, numerical results indicated that the spectral weight in the $\psi\psi^*$ 
continuum is rather small except near the edges, and there exists a large fraction 
of spectral weights above the continuum \cite{Lefmann,Muller}. 
\par
In this Letter, mainly focusing attention on behaviors of $S^{+-}(k,\omega)$ in magnetic fields, 
we identify excitations having large spectral weights to clarify overall dynamical features of 
the Heisenberg chain in magnetic fields. 
\par
\begin{figure}
\includegraphics[width=6.5cm]{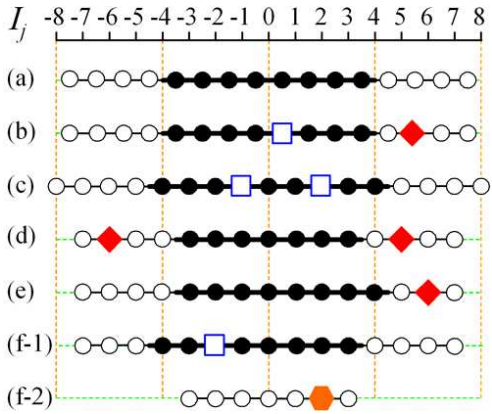}
\caption{Distributions of Bethe quantum numbers \{$I_j$\}. Filled symbols denote \{$I_j$\} of the ground state of $M$=8 
and typical excited states in dynamical structure factors for 0$\le$$k$$\le$$\pi$ in $L$=24. 
Blue open square and red solid diamond denote psinon ($\psi$) and 
antipsinon ($\psi^*$) \cite{Karbach_psinon}, which are located at points on thick and thin lines, respectively. 
(a) Ground state. 
(b) $\psi\psi^*$ for $S^{zz}(k,\omega)$. 
(c) 2$\psi$ for $S^{-+}(k,\omega)$. 
(d) 2$\psi^*$ for the continuum near $k$=$\pi$ in $S^{+-}(k,\omega)$. 
(e) 1$\psi^*$ for the low-energy mode near $k$=0 in $S^{+-}(k,\omega)$. 
(f-1) 1$\psi$ of \{$I_j^{\rm r}$\} for 2-string solutions in $S^{+-}(k,\omega)$. 
(f-2) $I_1^{\rm i}$ for the $n$-string of S$n$ ($n$$\ge$2) in $S^{+-}(k,\omega)$.}
\label{fig:Ij}
\end{figure}

We consider the spin-1/2 antiferromagnetic Heisenberg chain with $L$ sites, $M$ down spins 
($M$$\le$$L$/2) and periodic boundary conditions. The Hamiltonian is defined as 
\begin{equation}
{\cal H}=J\sum_{x=1}^L{\mbox {\boldmath $S$}}_{x}\cdot{\mbox {\boldmath $S$}}_{x+1}-HS^z, 
\label{eq:Ham}
\end{equation}
where ${\mbox {\boldmath $S$}}_{x}$ is the spin-1/2 operator at site $x$, and $J$$>$0. 
Magnetic field $H$ at $S^z/L$=1/2$-$$M/L$ in the thermodynamic limit is 
obtained in Ref.~\cite{Griffiths}. 
In the Bethe ansatz \cite{BetheAnsatz}, the wave function, energy and momentum 
of an eigenstate are expressed by a set of rapidities \{$\Lambda_j$\} which is obtained from 
the Bethe equation: 
$L\arctan$($\Lambda_j$)=$\pi$$I_j$$+$$\sum_{l=1}^M$$\arctan$[($\Lambda_j$$-$$\Lambda_l$)/2], 
once a set of \{$I_j$\} is given. 
Here, $I_j$ ($j$=1$\sim$$M$) are called Bethe quantum numbers, which are integers (half-odd integers) 
within $|I_j|$$\le$($L$$-$$M$$-$1)/2 for odd (even) $L$$-$$M$ as in Fig.~\ref{fig:Ij} (a-e) for solutions of real \{$\Lambda_j$\}. 
Distributions of \{$I_j$\} are somewhat analogous to momentum distributions of spinless fermions. 
As in Fig.~\ref{fig:Ij} (b), a hole (particle) created inside (outside) the \{$I_j$\} of the ground state 
is called psinon (antipsinon) and denoted by $\psi$ ($\psi^*$) \cite{Karbach_Szz,Karbach_psinon}. 
\par
It is also known that there are solutions with complex rapidities \cite{BetheAnsatz}. 
Later, we will consider solutions with an $n$-string ($n$$\ge$2), i.e. a set of $n$ rapidities expressed as 
$\Lambda_j$=${\bar \Lambda}$$+$$\i$($n$$+$1$-$2$j$)$+$$\i$$\eta_j$ for $j$=1$\sim$$n$, 
where ${\bar \Lambda}$ is real, and $\eta_j$=$O$(e$^{-c L}$) with $c$$>$0 \cite{BetheAnsatz,Takahashi,Caux,Hagemans}. 
We take real $\Lambda_j$ for $j$$\ge$$n$$+$1, and denote the $n$-string solutions 
by S$n$. These solutions are specified by two sets of \{$I_j$\} \cite{Takahashi}: One is for the real rapidities, 
and the other is for the $n$-string, which we denote by \{$I_j^{\rm r}$\} and \{$I_j^{\rm i}$\}, 
respectively. For S$n$ ($n$$\ge$2), \{$I_j^{\rm i}$\}=$I_1^{\rm i}$, $|I_1^{\rm i}|$$\le$$L$/2$-$$M$ as in Fig.~\ref{fig:Ij}~(f-2). 
We denote $\psi$ and $\psi^*$ of \{$I_j^{\rm r}$\} by $\psi_{\rm s}$ and $\psi_{\rm s}^*$, respectively, and 
regard \{$I_j^{\rm r}$\} without $\psi_{\rm s}$ or $\psi_{\rm s}^*$ also as a part of those with $\psi_s\psi_s^*$. 
\par
We investigate behaviors of dynamical structure factors defined as 
$S^{{\bar \alpha}\alpha}(k,\omega)$=$\sum_iM^{{\bar \alpha}\alpha}(k,\epsilon_k^i)\delta(\omega-\epsilon_k^i)$ 
for $\alpha$$=$$-$, $+$ and $z$. Here, $M^{{\bar \alpha}\alpha}(k,\epsilon_k^i)$ is 
the transition rate between the ground state $|{\rm G.S.}\rangle$ in a magnetic field 
and an excited state $|k,\epsilon_k^i\rangle$ with excitation energy $\epsilon_k^i$ 
and momentum $k$, defined as 
$M^{{\bar \alpha}\alpha}(k,\epsilon_k^i)$=$|\langle k,\epsilon_k^i|S^\alpha_{k}|{\rm G.S.}\rangle|^2$. 
We calculated $M^{{\bar \alpha}\alpha}(k,\epsilon_k^i)$ of the Heisenberg chain, following 
Refs.~\cite{Kitanine,Biegel,BiegelXXZ,Hagemans,Caux,Sato} where $M^{{\bar \alpha}\alpha}(k,\epsilon_k^i)$ 
is expressed in a determinant form whose matrix elements are expressed in terms of rapidities \cite{Kitanine}. 
As for string solutions with an $n$-string ($n$$\ge$2), we calculated 
$M^{{\bar \alpha}\alpha}(k,\epsilon_k^i)$ after transforming the matrices so that singularities in determinants 
can be cancelled as in Refs. \cite{Caux,Hagemans}. 
In $S^{+-}(k$=$0,\omega)$, we took into account the resonant mode 
of the field-induced magnetization \cite{Muller,Karbach_psinon,Biegel,Nishimoto,Lefmann}. 
In this Letter, we show results on dynamical structure factors in units of $1/J$ for 0$\le$$k$$\le$$\pi$, 
noting $S^{{\bar \alpha}\alpha}(k,\omega)$=$S^{{\bar \alpha}\alpha}(-k,\omega)$, $\alpha$$=$$-$, $+$ and $z$. 
\par
\begin{figure}
\includegraphics[width=8.6cm]{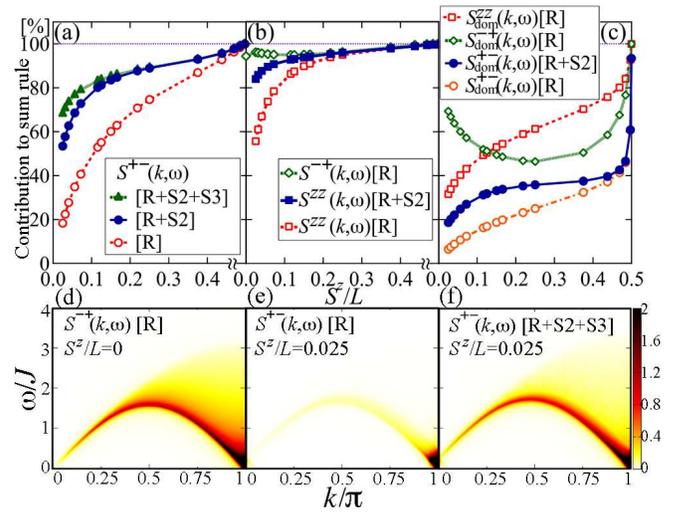}
\caption{(a,b) Contributions to sum rules from solutions of up to 2$\psi$2$\psi^*$, 
string solutions with a 2-string and those with a 3-string, denoted by R, S2 and S3, respectively, 
in $L$=320 for (a) $S^{+-}(k,\omega)$, (b) $S^{-+}(k,\omega)$ and $S^{zz}(k,\omega)$. 
(c) Contributions of dynamically dominant excitations in $L$=2240. 
(d) $S^{-+}(k,\omega)$ in zero field. (e,f) $S^{+-}(k,\omega)$ at $S^z/L$=0.025 
without and with the contributions of S2 and S3. In (d-f), the data obtained in $L$=320 
are broadened in a Lorentzian form with full-width at half maximum (FWHM) $0.08J$. 
For S2 and S3 in (a,b,f), we took the $O(L^3)$ states described in the text.}
\label{fig:sumrule}
\end{figure}
We calculated the spectral weight in $L$=320, using up to 2$\psi$2$\psi^*$ excitations, 
and evaluated the contributions to sum rules \cite{totalweight}. 
The results are shown by open symbols in Fig.~\ref{fig:sumrule}~(a) and (b). 
The spectral weight of up to 2$\psi$2$\psi^*$ excitations in $S^{-+}(k,\omega)$ satisfies 
more than 90\% of the sum rule in $L$=320 as shown by open diamonds in Fig.~\ref{fig:sumrule}~(b). 
The spectral weight in $S^{+-}(k,\omega)$ severely decreases in the small $S^z$ regime 
as shown by open circles in Fig.~\ref{fig:sumrule}~(a). Figure~\ref{fig:sumrule}~(e) shows 
$S^{+-}(k,\omega)$ in the small $S^z$ regime. Although $S^{+-}(k,\omega)$ 
should be continuously connected to that in zero field (Fig.~\ref{fig:sumrule}~(d)) 
as the magnetic field decreases, most of the spectral weight in $S^{+-}(k,\omega)$ is missing. 
In particular, the large intensity near the des Cloizeaux-Pearson mode (dCP) \cite{dCP} 
(the lower edge of the continuum in zero field) is almost lost. This implies that there are 
other important excitations for $S^{+-}(k,\omega)$. 
\par
To explain the origin of the missing spectral weight, we consider S2 and S3 defined above. 
Among many states in S2 and S3, we found that S2 with $\psi_{\rm s}\psi_{\rm s}^*$ 
and S3 with 2$\psi_{\rm s}$ have large weights in $S^{+-}(k,\omega)$. 
By taking into account these string solutions, $S^{+-}(k,\omega)$ in the small $S^z$ regime almost recovers 
the missing weight near the dCP mode and that of the continuum near $k$=$\pi$ 
as shown in Fig.~\ref{fig:sumrule}~(f). The intensity near the dCP mode is mainly 
due to the S2, and that of the continuum near $k$=$\pi$ is mainly due to the S3. 
The contributions from these string solutions increase as the magnetic field decreases 
as shown by solid symbols in Fig.~\ref{fig:sumrule}~(a). By taking them into account, 
more than 80\% of the total spectral weight in $S^{+-}(k,\omega)$ 
is explained in a wide range of $S^z$ in $L$=320~\cite{size_effect}. 
\par
As for $n$-string solutions ($n$$\ge$2) in $S^{zz}(k,\omega)$, we found that a large spectral weight 
is carried by S2 with 2$\psi_{\rm s}$ in low fields. As shown by solid squares 
in Fig.~\ref{fig:sumrule}~(b), the decrease of $S^{zz}(k,\omega)$ of up to 2$\psi$2$\psi^*$ 
excitations in the small $S^z$ regime (open squares in Fig.~\ref{fig:sumrule}~(b)) is almost 
compensated for by the S2, and more than 90\% of the total spectral weight in $S^{zz}(k,\omega)$ 
is explained in a wide range of $S^z$ in $L$=320~\cite{size_effect}. 
These string solutions are relevant to the continuum near $k$=$\pi$ in low fields. 
\par
\begin{figure}
\includegraphics[width=8.6cm]{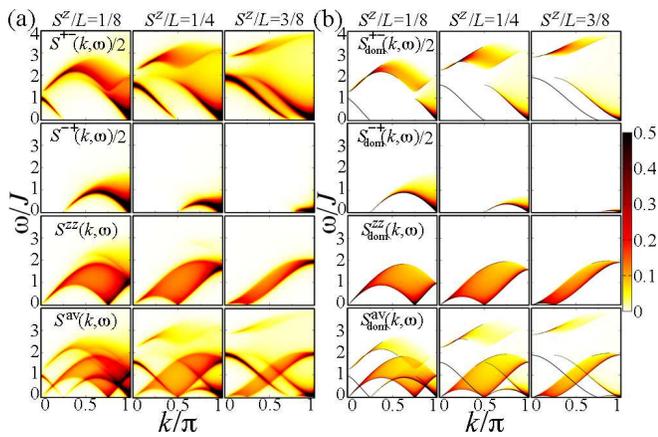}
\caption{(a)$\!$ $S^{+-}(k,\omega)$/2, $S^{-+}(k,\omega)$/2, 
$S^{zz}(k,\omega)$ and $S^{\rm av}(k,\omega)$ (from above) at $S^z/L$=1/8, 1/4 and 3/8 (from the left) 
in $L$=320. The data are broadened in a Lorentzian form 
with FWHM=0.08$J$. (b) Same as (a) but those of dynamically dominant excitations in $L$=2240 with the product ansatz.}
\label{fig:Mdep}
\end{figure}
Figure~\ref{fig:Mdep}~(a) shows behaviors of dynamical structure factors in magnetic fields 
in $L$=320, where we took into account excitations of up to 2$\psi$2$\psi^*$ and 
the above-described string solutions. 
Hereafter, we use these excitations to calculate dynamical structure factors. 
In $S^{+-}(k,\omega)$, there are three sets of dominant continua as shown in the top row 
of Fig.~\ref{fig:Mdep}~(a). The high-energy continuum ($\omega/J$$\agt$2), which is mainly 
due to S2, goes up to higher energies as the magnetic field increases, separated from 
the low-energy continua. 
Although this continuum has been observed in the high-field regime by numerical calculations 
\cite{Muller,Nishimoto,Lefmann}, the relation to the S2 has not been clarified. 
Near the saturation field, these string solutions reduce to bound states of two overturned spins 
from the ferromagnetic state \cite{BetheAnsatz,Muller} with the excitation energy given by 
$\omega(k)$=$\frac{J}{2}$(3$-$$\cos k$)+$H$. 
As for $S^{-+}(k,\omega)$ 
and $S^{zz}(k,\omega)$, there is basically one continuum for each correlation as shown 
in the second and third rows of Fig.~\ref{fig:Mdep}~(a), respectively. The additional high-energy 
continuum at $S^z/L$=1/8 in $S^{zz}(k,\omega)$ near $k$=$\pi$ mainly comes from S2. 
\par
From many excitations considered above, we extract dynamically dominant 
excitations of $O(L^2)$ states which characterize the behaviors 
of Fig.~\ref{fig:Mdep}~(a). Then, the product ansatz \cite{Karbach_Szz}: 
$S^{{\bar \alpha}\alpha}(k,\omega)$=$M^{{\bar \alpha}\alpha}(k,\omega)D(k,\omega)$ 
with $D(k,\epsilon_k^i)$=$2/(\epsilon_k^{i+1}$$-$$\epsilon_k^{i-1})$ can be applied. 
Here, in energy regions where there is more than one sequence of states, we took 
the one with the largest spectral weight. 
Hereafter, we denote $S^{{\bar \alpha}\alpha}(k,\omega)$ of the dynamically dominant excitations 
as $S^{{\bar \alpha}\alpha}_{\rm dom}(k,\omega)$ for $\alpha$=$-$, $+$ and $z$. 
Their contributions to sum rules in $L$=2240 are shown in Fig.~\ref{fig:sumrule}~(c). 
\par
The top row of Fig.~\ref{fig:Mdep}~(b) shows $S^{+-}_{\rm dom}(k,\omega)$. 
The high-energy continuum near $\omega/J$$\agt$2 is due to S2 
with 1$\psi_{\rm s}$ as in Fig.~\ref{fig:Ij}~(f-1), which can be regarded 
as a part of the above-mentioned S2 with $\psi_{\rm s}\psi_{\rm s}^*$. 
The low-energy continuum near $k$=$\pi$ comes from 2$\psi^*$ excitations 
where two $\psi^*$s are located on opposite sides of the filled region of \{$I_j$\} 
from each other as in Fig.~\ref{fig:Ij}~(d). The low-energy mode near $k$=0 is due to 
excitations with one $\psi^*$ in the right empty region of \{$I_j$\} 
for 0$\le$$k$$\le$$\pi$ as in Fig. \ref{fig:Ij} (e). This mode has been mentioned 
in the literature~\cite{Muller,Karbach_psinon,Biegel,Nishimoto,Lefmann}. 
\par
The dynamically dominant excitations for $S^{-+}(k,\omega)$ and $S^{zz}(k,\omega)$ are known 
as 2$\psi$ and $\psi\psi^*$ excitations as in Fig.~\ref{fig:Ij}~(c) and (b), 
respectively \cite{Karbach_Szz,Karbach_psinon}. 
The results are shown 
in the second and third rows of Fig.~\ref{fig:Mdep}~(b). 
\par
Noting that 2$\psi$, $\psi\psi^*$ and 2$\psi^*$ excitations are dynamically dominant 
in $S^{-+}(k,\omega)$, $S^{zz}(k,\omega)$ and $S^{+-}(k,\omega)$ and that these excitations 
have $S^z$=$+1$, 0 and $-1$, respectively, we can naturally assign $S^z$=$+$1/2 and $-$1/2 
to $\psi$ and $\psi^*$, respectively. Also, noting that excitations of S2 with $\psi_{\rm s}\psi_{\rm s}^*$ 
and those with 2$\psi_{\rm s}$ have large weights in $S^{+-}(k,\omega)$ and $S^{zz}(k,\omega)$ and 
that these excitations have $S^z$=$-$1 and 0, respectively, 
we can naturally interpret the quasiparticle representing the 2-string (Fig.~\ref{fig:Ij}~(f-2)) 
as a bound state of two $\psi^*$s which carries $S^z$=$-$1. This assignment is also 
applicable to 4-spinon states of $S^{-+}(k,\omega)$ in zero field which have four spinons and a 2-string. 
The 1$\psi^*$ mode near $k$=0 in $S^{+-}(k,\omega)$ can also be regarded 
as the most dominant part of a 2$\psi^*$ continuum where both $\psi^*$s are located 
in the right empty region of \{$I_j$\} (Fig.~\ref{fig:Ij}~(e) as compared with (d)) for 0$\le$$k$$\le$$\pi$. 
But, since the spectral weight in this continuum is very small except the 1$\psi^*$ mode, 
the $\psi^*$ in this mode practically behaves as a quasiparticle carrying $S^z$=$-$1, 
which reduces to a magnon carrying $S^z$=$-$1 above the saturation field. 
\par
\begin{figure}
\includegraphics[width=8.6cm]{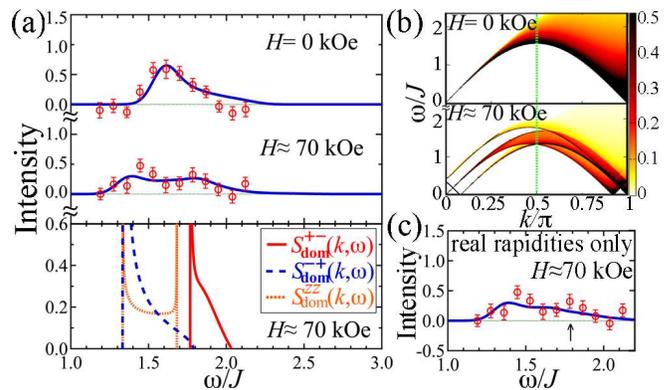}
\caption{Comparison with experimental results on CuCl$_2\cdot$2N(C$_5$D$_5$). 
(a) Solid lines in upper two panels are the present results of $S^{\rm av}(k,\omega)$ 
at $k$=$\pi$/2 in $L$=320. 
The data are broadened in a gaussian form with FWHM$\simeq$0.17$J$. 
Open symbols are experimental results in Ref.~\cite{CPC}. 
The height is rescaled after subtracting the background (=180 counts). 
The lowest panel shows lineshapes of $S^{{\bar \alpha}\alpha}_{\rm dom}(k,\omega)$ 
($\alpha$=$-$, $+$ and $z$) at $k$=$\pi$/2 in $L$=2240. 
(b) $S^{\rm av}_{\rm dom}(k,\omega)$ in $L$=2240. 
(c) Same as the middle panel of (a) except that the solid line is obtained from solutions with real rapidities only. 
The arrow indicates the high-energy peak observed in the experiment~\cite{CPC}.}
\label{fig:CPC}
\end{figure}
To confirm the relevance to real materials, we compare the present 
results with available experimental data on quasi-one-dimensional materials 
which can be effectively regarded as spin-1/2 antiferromagnetic Heisenberg chains. 
In inelastic neutron scattering experiments, a quantity proportional 
to $S^{\rm av}(k,\omega)$ is observed. Here, we define 
$S^{\rm av}(k,\omega)$$\equiv$$[S^{-+}(k,\omega)$+$S^{+-}(k,\omega)$+$4S^{zz}(k,\omega)]/6$. 
The fourth row of Fig.~\ref{fig:Mdep} shows the results of $S^{\rm av}(k,\omega)$ and $S^{\rm av}_{\rm dom}(k,\omega)$, 
where $S^{\rm av}_{\rm dom}(k,\omega)$ denotes $S^{\rm av}(k,\omega)$ of the dynamically dominant excitations. 
\par
In Fig.~\ref{fig:CPC}~(a), we compare the present results with experimental results on 
CuCl$_2\cdot$2N(C$_5$D$_5$) (CPC) \cite{CPC}. We broadened the numerical results 
of $S^{\rm av}(k,\omega)$ in a gaussian form in zero field, and rescaled the experimental data 
in Ref.~\cite{CPC} after subtracting the background. Using the same broadening and 
rescaling parameters, we calculated $S^{\rm av}(k,\omega)$ at $H$$\approx$70 kOe and 
compared it with experimental results, where we used $g$-factor $g$=2.08 and 
$J$=27.32 K \cite{CPCprm}. 
The positions and intensities of the two peaks at $H$=70 kOe observed in the experiment are 
reasonably reproduced by the present results as shown in the middle panel of Fig.~\ref{fig:CPC}~(a). 
The high- and low-energy peaks are mainly due to 
S2 in $S^{+-}(k,\omega)$ and 2$\psi$ excitations in $S^{-+}(k,\omega)$, 
respectively, as shown in the lowest panels of Fig.~\ref{fig:CPC}~(a) and (b). 
For comparison, the results of solutions with real rapidities only 
are shown in Fig.~\ref{fig:CPC}~(c). Obviously, the spectral weight at the high-energy peak 
is missing, if $n$-string solutions ($n$$\ge$2) are neglected. 
\par
\begin{figure}
\includegraphics[width=8.6cm]{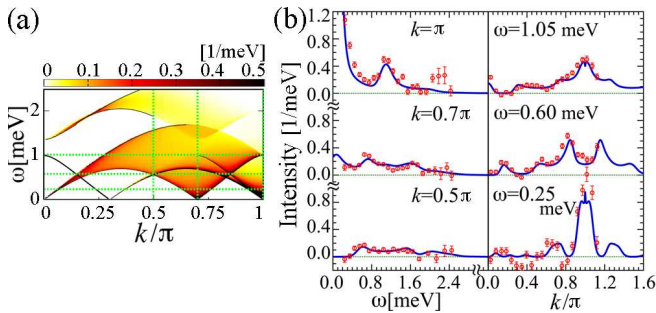}
\caption{Comparison with experimental results on Cu(C$_4$H$_4$N$_2$)(NO$_3$)$_2$. 
(a) $S^{\rm av}_{\rm dom}(k,\omega)$ at $S^z/L$=0.15 in $L$=2240. 
(b) Solid blue lines denote $S^{\rm av}(k,\omega)$ in $L$=320 along green dotted lines in (a). 
The data are broadened in a gaussian form with standard deviation $\sigma$=0.1 meV. 
Open symbols are experimental results of intensity at $H$=8.7 T in Ref.~\cite{CuPzN}. 
Here, we subtracted the background equal to that of $H$=0~\cite{CuPzN}.}
\label{fig:CuPzN}
\end{figure}
Figure~\ref{fig:CuPzN}~(b) shows comparisons with experimental results on 
Cu(C$_4$H$_4$N$_2$)(NO$_3$)$_2$ (CuPzN) \cite{CuPzN} along scans denoted 
by green dotted lines in Fig.~\ref{fig:CuPzN}~(a). The present results reasonably agree with the experimental results. 
Figure~\ref{fig:CuPzN}~(a) is plotted so that it can be easily compared with Fig.~4~(a) 
in Ref.~\cite{CuPzN}. A signature of the high-energy continuum originating from the 2-string solutions was 
observed near $k$$\simeq$0.3$\pi$ and $\omega$$\simeq$2 meV in the experiment~\cite{CuPzN}. 
\par
Signatures of the 2-string solutions in an anisotropic two-dimensional system 
will be shown elsewhere \cite{q1DH}. 
\par
In summary, we have investigated dynamical properties of the spin-1/2 antiferromagnetic 
Heisenberg chain in magnetic fields, using Bethe-ansatz solutions. We found 
that string solutions with a 2-string have considerable spectral weight in $S^{+-}(k,\omega)$, 
and can be regarded as dynamically dominant excitations of $S^{+-}(k,\omega)$. 
This indicates that not only $\psi$ and $\psi^*$ but also the quasiparticle representing 
the 2-string plays an important role for the dynamical properties in magnetic fields. 
The continuum of these solutions continuously connects the dCP mode in the zero-field limit and 
that of bound states of overturned spins from the ferromagnetic state \cite{BetheAnsatz} 
near the saturation field. 
Another finding is that 2$\psi^*$ excitations are identified as the dominant excitations 
of $S^{+-}(k,\omega)$ near $k$=$\pi$ in the low-energy regime, which leads to 
a natural interpretation of $\psi$ and $\psi^*$ as quasiparticles 
in magnetic fields carrying $S^z$=$+$1/2 and $-$1/2, respectively. 
Comparisons with available experimental results reasonably support the relevance 
to real materials. For clear identification of dynamically dominant excitations, 
further experiments in higher fields at high energies are desired. 
Behaviors shown in this Letter are expected to be more or less true 
for general spin-1/2 XXZ chains in magnetic fields. 
\par
{\it Acknowledgements -}
I am grateful to L. Balents, M. Shiroishi, M. Arikawa, O.A. Starykh, R. Coldea, M. Takahashi 
and A. Tanaka for discussions, helpful comments and suggestions. This work was supported 
by KAKENHI 20740206 and 20046015, and World Premier International Research 
Center Initiative, MEXT, Japan.


\begin{references}
\bibitem{BetheAnsatz}H. Bethe, Z. Phys. {\bf 71}, 205 (1931).
\bibitem{FT}L. D. Faddeev {\it et al.}, Phys. Lett. A {\bf 85}, 375 (1981).
\bibitem{HSspinon}F. D. M. Haldane, Phys. Rev. Lett. {\bf 66}, 1529 (1991).
\bibitem{KCuF3}D. A. Tennant {\it et al.}, Phys. Rev. B {\bf 52}, 13368 (1995).
\bibitem{CuBenz}D. C. Dender {\it et al.}, Phys. Rev. B {\bf 53}, 2583 (1996).
\bibitem{CuPzN0}P. R. Hammar {\it et al.}, Phys. Rev. B {\bf 59}, 1008 (1999).
\bibitem{Ishimaru}N. Ishimaru, and H. Shiba, Prog. Theor. Phys. {\bf 57}, 1862 (1977); 
{\it ibid.} {\bf 64}, 479 (1980).
\bibitem{CuPzN}M. B. Stone {\it et al.}, Phys. Rev. Lett. {\bf 91}, 037205 (2003).
\bibitem{Muller}G. M\"uller {\it et al.}, Phys. Rev. B {\bf 24}, 1429 (1981). 
\bibitem{Karbach_psinon}M. Karbach  {\it et al.}, Phys. Rev. B {\bf 66}, 054405 (2002).
\bibitem{Karbach_Szz}M. Karbach {\it et al.}, Phys. Rev. B {\bf 62}, 14871 (2000).
\bibitem{Fledderjohann}A. Fledderjohann {\it et al.}, Phys. Rev. B {\bf 54}, 7168 (1996). 
\bibitem{CauxPRL}J. -S. Caux {\it et al.}, Phys. Rev. Lett. {\bf 95}, 077201 (2005). 
\bibitem{CauxAffleck}R. G. Pereira {\it et al.}, Phys. Rev. Lett. {\bf 96}, 257202 (2006). 
\bibitem{Sato}J. Sato {\it et al.}, J. Phys. Soc. Jpn.  {\bf 73}, 3008 (2004). 
\bibitem{Caux}J. -S. Caux {\it et al.}, J. Stat. Mech. P09003 (2005).
\bibitem{BiegelXXZ}D. Biegel {\it et al.}, J. Phys. A: Math. Gen. {\bf 36} 5361 (2003).
\bibitem{Biegel}D. Biegel {\it et al.}, Europhys. Lett. {\bf 59}, 882 (2002). 
\bibitem{Nishimoto}S. Nishimoto {\it et al.}, Int. J. Mod. Phys. B {\bf 21}, 2262 (2007). 
\bibitem{Lefmann}K. Lefmann {\it et al.}, Phys. Rev. B {\bf 54}, 6340 (1996). 
\bibitem{Griffiths}R. B. Griffiths, Phys. Rev. {\bf 133}, A768 (1964).
\bibitem{Takahashi}M. Takahashi, Prog. Thoer. Phys. {\bf 46}, 401 (1971).
\bibitem{Hagemans}R. Hagemans {\it et al.}, AIP Conf. Proc. {\bf 846}, 245 (2006). 
\bibitem{Kitanine}N. Kitanine {\it et al.}, Nucl. Phys. B {\bf 554}, 647 (1999).
\bibitem{totalweight}The total spectral weights of 
$S^{{\bar \alpha}\alpha}(k,\omega)$ ($\alpha$=$-$,$+$ and $z$) are $L$$-$$M$, $M$ and 
$L/4$$-$$(S^{z})^2/L$, respectively \cite{Caux,CauxPRL}. 
\bibitem{dCP}J. des Cloizeaux {\it et al.}, Phys. Rev. {\bf 128}, 2131 (1962).
\bibitem{size_effect}Judging from small size-dependence in $L$$\alt$320 for Fig. \ref{fig:sumrule} (a,b), 
we believe the physics will hold for larger systems. 
\bibitem{CPC}I. U. Heilmann {\it et al.}, Phys. Rev. B {\bf 18}, 3530 (1978).
\bibitem{CPCprm}J. A. Chakhalian {\it et al.}, Phys. Rev. Lett. {\bf 91}, 027202 (2003). 
\bibitem{q1DH}M. Kohno (unpublished).
\end{references}

\end{document}